\documentclass[11pt]{article}
\usepackage[utf8]{inputenc}
\pdfoutput=1
\usepackage{amssymb,amsmath,mathrsfs,enumerate}
\numberwithin{equation}{section}
\usepackage{mathtools}
\usepackage{graphicx,rotate,multicol}
\usepackage{float}
\usepackage{tocloft}
\usepackage{subcaption}
\usepackage[margin=10pt,labelfont=bf]{caption}
\usepackage{cite}
\usepackage{soul}
\usepackage[dvipsnames]{xcolor}
\usepackage[colorlinks=true,
            linkcolor=blue,
            urlcolor=NavyBlue,
            citecolor=ForestGreen]{hyperref}
\usepackage{multirow}
\usepackage{array}
\usepackage{placeins}
\usepackage{slashed}

\makeatletter
\let\Hy@linktoc\Hy@linktoc@page
\makeatother

\usepackage{color}
\definecolor{ourcolor}{rgb}{0.7, 0.25, 0.05}
\usepackage{tikz,braket}
\long\def\rpl#1!!#2!!{\textcolor{red}{#1} \textcolor{blue}{#2}}

\let\bar=\overline

\def \order(#1){{\mathcal O} \left(#1 \right)}

\evensidemargin=\oddsidemargin
\allowdisplaybreaks[4]

\title{\color{black}{\bf Neutrinos from captured dark matter in galactic stars}}

\author {{\bf Debajit Bose,}\footnote{\href{mailto:debajitbose550@gmail.com}{debajitbose550@gmail.com}}
\hspace{4pt}  {\bf Rohan Pramanick,}\footnote{\href{mailto:rohanpramanick25@gmail.com}{rohanpramanick25@gmail.com}}
{\bf \ and Tirtha Sankar Ray}\,\footnote{\href{mailto:tirthasankar.ray@gmail.com}{tirthasankar.ray@gmail.com}}  
\\[5pt]
{\it Department of Physics, Indian Institute of Technology Kharagpur, Kharagpur 721302, India}}

\date{}

\begin{document}

\maketitle

\begin{abstract}
Sub-GeV neutrinos produced in a stellar core may emerge from main sequence stars, white dwarfs and brown dwarfs producing possible observable signals of dark matter capture. A distribution of these stars near the Milky Way galactic center will produce a neutrino flux that can be probed at Earth based neutrino observatories like Super-Kamiokande and Hyper-Kamiokande. We demonstrate that this can provide a handle to probe dark matter masses in the $200\,$MeV\,$-$\,$2\,$GeV mass scales that compares favourably with present day direct detection bounds.
\end{abstract}
%
%
%%%%%%%%%%%%%%%%%%%%%%%%%%%%%%%%%%%%%%%%%%%%%%%%%%%%%%%%%%%%%%%%%%%%%%%%%%%%%%%%%%%%%%%%%%%%
\section{Introduction}
\label{sec:intro}
%%%%%%%%%%%%%%%%%%%%%%%%%%%%%%%%%%%%%%%%%%%%%%%%%%%%%%%%%%%%%%%%%%%%%%%%%%%%%%%%%%%%%%%%%%%%

Dark matter (DM) with a portal to the visible sector can be captured by the celestial stars by gravitational focusing followed by dissipative scatterings \cite{Press1985CaptureBT,Gould:1987ju,Gould:1987ir,Bramante:2023djs}. Once captured  the  dark matter particles  settle towards the stellar core where their density  rises facilitating annihilations to visible sector states. The density and ionisation of the stellar interior impedes the emergence of the standard model (SM) particles with the exception of low energetic neutrinos. Notwithstanding the rising  scattering cross-section with energy, we find that sub-GeV neutrinos may emerge from the classical main sequence stars and the relatively less dense compact stars like the brown dwarfs and white dwarfs. These neutrinos provide a natural signature of dark matter capture in these stars.

However, the  attenuation encountered in propagation within the stars and low detection probabilities of neutrinos on Earth based observatories imply that signals from individual stars remain a challenging proposition. A more optimistic possibility for neutrino signals of captured dark matter arises from the distribution of stars that are clustered near the galactic center. In this article,  we study the sensitivity of such signals to probe sub-GeV scale dark matter at neutrino observatories like Super-Kamiokande (SK) \cite{Super-Kamiokande:2002weg} and the proposed Hyper-Kamiokande (HK) \cite{Hyper-Kamiokande:2018ofw}.
%
%
%%%%%%%%%%%%%%%%%%%%%%%%%%%%%%%%%%%%%%%%%%%%%%%%%%%%%%%%%%%%%%%%%%%%%%%%%%%%%%%%%%%%%%%%%%%%
\begin{figure}[t]
\centering
\includegraphics[width=0.5\linewidth]{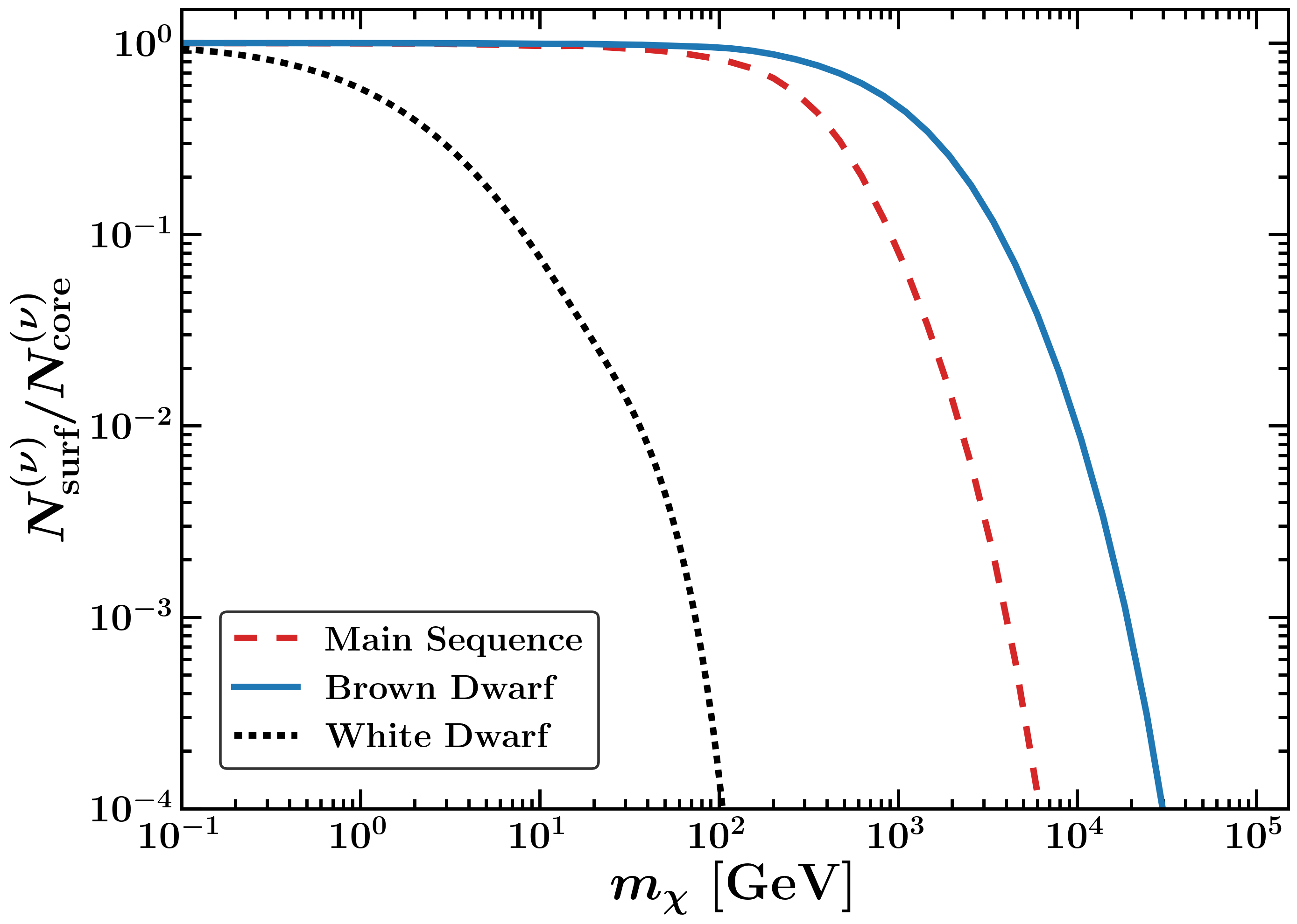}
\caption{The ratio between the propagated neutrino flux at the surface and the produced neutrino flux at the core for MS, BD, and WD stars represented by the red (dashed), blue (solid), and black (dotted) lines respectively.}
\label{fig:Neutrino_flux_wo_Med}
\end{figure}
%%%%%%%%%%%%%%%%%%%%%%%%%%%%%%%%%%%%%%%%%%%%%%%%%%%%%%%%%%%%%%%%%%%%%%%%%%%%%%%%%%%%%%%%%%%%
%
%
%%%%%%%%%%%%%%%%%%%%%%%%%%%%%%%%%%%%%%%%%%%%%%%%%%%%%%%%%%%%%%%%%%%%%%%%%%%%%%%%%%%%%%%%%%%%
\section{Neutrinos from stellar core}
\label{sec:neutrinos_from_stars}
%%%%%%%%%%%%%%%%%%%%%%%%%%%%%%%%%%%%%%%%%%%%%%%%%%%%%%%%%%%%%%%%%%%%%%%%%%%%%%%%%%%%%%%%%%%%

Dark matter capture by stars in our galaxy reaches its crescendo near the galactic center where both the dark matter density and the stellar distribution profiles are at their peak. The main sequence (MS) stars, white dwarfs (WD) and brown dwarfs (BD) are more populous than the neutron stars (NS) while the latter remains relevant owing to their high density and compact structure. If the dark sector is not asymmetric in nature, then the captured DM particles can self-annihilate after being accumulated inside the stellar core. The DM may directly annihilate to neutrinos or they may be produced down stream in a cascade eventually emerging from the stellar interior. It is unlikely that other visible sector particles can escape from the dense ionised interiors and usually end up dissipating their energy inside the stars \cite{Kouvaris:2007ay,McCullough:2010ai,Raj:2017wrv,Baryakhtar:2017dbj,Joglekar:2019vzy,Bell:2020jou,Leane:2020wob,Maity:2021fxw,Bell:2021fye}. In this work, we adopt a model agnostic paradigm to constrain the parameter space of captured DM that primarily annihilate to neutrinos. Underlying UV complete realizations may be readily constructed by incorporating a massive mediator that has significant couplings to neutrinos and suppressed branching to SM quarks and leptons \cite{Lin:2022hnt}. Here we study the possibility of neutrinos produced at the core emerging from the surface of the stars as a signature of DM capture \cite{Krauss1985,Srednicki1987,Belotsky:2002sv,Belotsky:2008vh,Zentner:2009is,Bernal:2012qh,IceCube:2016dgk,IceCube:2021xzo,Bell:2021esh,Maity:2023rez,Bose:2023yll}. Due to their weak interactions with other SM constituents, the neutrinos undergo scattering with stellar particles while propagating within the stellar atmosphere. The propagation within the star is affected by scattering, tau regeneration, and medium oscillations. In Fig. \ref{fig:Neutrino_flux_wo_Med}, we explore the impact of this  propagation within stellar atmospheres across three types of stars of interest using \texttt{nuSQuIDS} \cite{Arguelles:2021twb}. The ratio of the attenuated neutrino flux at the surface $\left( N^{(\nu)}_{\rm surf} \right)$ to that  produced at the core $\left( N^{(\nu)}_{\rm core} \right)$ for a given DM mass $(m_\chi)$ is obtained by considering the high energy tail of the differential flux. We integrate over the energy bin $\left[ m_\chi/5 , m_\chi \right]$ considering the spectra of DM annihilation to neutrinos peaks near the DM mass. Consequently, the most optimistic bounds are expected to arise from the high-energy tail of the neutrino flux.

As can be seen from Fig. \ref{fig:Neutrino_flux_wo_Med}, for the more dense stellar environment, it becomes increasingly difficult for the higher energetic neutrinos to escape from the stars. Within the energy range under consideration, for NS, we find that neutrinos produced within the core through captured DM annihilation, are unable to escape the stars sufficiently to produce any observable effect. So, for NS, the only viable approach to obtain an observable annihilation signature is through the annihilation into hidden sector particles, as explored in \cite{Leane:2021ihh,Bose:2021yhz}. For the WD, BD and MS stars, low energetic neutrinos can emerge from the stellar environment with an attenuated flux and can produce an observable signal. As depicted in Fig. \ref{fig:Neutrino_flux_wo_Med}, the observable neutrino signatures from these stars can extend into the TeV domain. However, due to heavy attenuation, the requisite cross-section for detectable signatures is currently ruled out by direct detection experiments. At the other end, capture of low mass DM is inhibited by evaporative effects \cite{Garani:2021feo} leaving  a narrow  window of DM mass between $200\,$MeV\,$-$\,$2\,$GeV, where the emergent neutrino signals remain relatively unattenuated that may be observable at Super-Kamiokande \cite{Super-Kamiokande:2002weg} or in the upcoming Hyper-Kamiokande detector \cite{Hyper-Kamiokande:2018ofw}.
%
%
%%%%%%%%%%%%%%%%%%%%%%%%%%%%%%%%%%%%%%%%%%%%%%%%%%%%%%%%%%%%%%%%%%%%%%%%%%%%%%%%%%%%%%%%%%%%
\begin{figure}[t]
\centering
\includegraphics[width=0.5\linewidth]{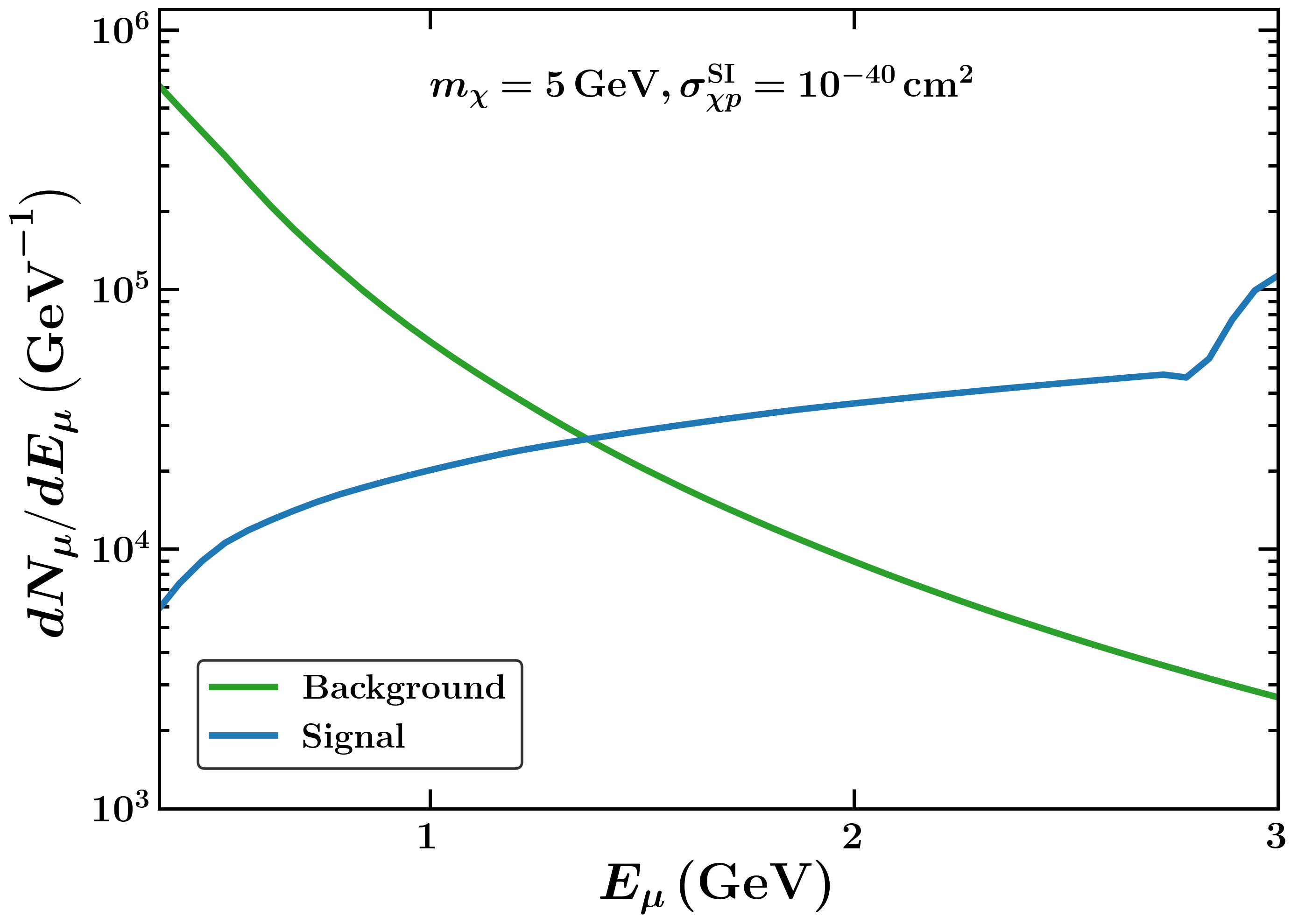}
\caption{{The differential muon event rates as given in Eq. \ref{eq:dN_mu-dE_mu} plotted with respect to muon energy for background (green) and signal events (blue) in the direct annihilation framework with DM mass of \(5\,\text{GeV}\) and spin-independent (SI) scattering cross-section of \(10^{-40} \, \text{cm}^2\), obtained by considering the WD distribution of stars in the galactic center.}}
\label{fig:Event_wo_Med_WD}
\end{figure}
%%%%%%%%%%%%%%%%%%%%%%%%%%%%%%%%%%%%%%%%%%%%%%%%%%%%%%%%%%%%%%%%%%%%%%%%%%%%%%%%%%%%%%%%%%%%
%
%
%%%%%%%%%%%%%%%%%%%%%%%%%%%%%%%%%%%%%%%%%%%%%%%%%%%%%%%%%%%%%%%%%%%%%%%%%%%%%%%%%%%%%%%%%%%%
\section{Dark matter capture at stellar distributions}
\label{sec:dm_capture}
%%%%%%%%%%%%%%%%%%%%%%%%%%%%%%%%%%%%%%%%%%%%%%%%%%%%%%%%%%%%%%%%%%%%%%%%%%%%%%%%%%%%%%%%%%%%

As a celestial object wanders in a galactic halo, it captures the ambient DM first by gravitationally focusing it in its direction and then through scattering processes that dissipates the incoming DM energy below the escape energy of the star. Thus the capture rate of DM by a star is a function of the DM density in the stellar neighbourhood and the DM scattering cross-section. The capture rate of DM particles within an astrophysical body can be schematically written as \cite{Bramante:2017xlb,Dasgupta:2019juq,Ilie:2020vec,Busoni:2017mhe,Garani:2017jcj}
\begin{gather}
\label{eq:Cap_rate}
C{_\star}(r) = \left( \dfrac{\rho_\chi (r)}{m_\chi} \right) \int_0^{R_{\star}} dr' \, 4 \pi r'^2 \int_0^{u_{\rm esc}} d_\chi \dfrac{f(u_\chi)}{u_\chi} \left( u_\chi^2 + v_{\rm esc}(r')^2 \right) \zeta(\sigma, u_\chi, r'),
\end{gather}
where $r'$ is the distance of the scattering region from the center of the star and $m_\chi$ is the mass of the DM particle and $\rho_\chi$ is the local DM density, $f \left( u_\chi \right)$ is the Maxwellian velocity distribution profile (for deviation see \cite{Bose:2022ola})  and $u_{\rm esc}$ be the  escape velocity of the halo DM. In Eq. \eqref{eq:Cap_rate} the function $\zeta(\sigma, u_\chi, r') $ captures the kinematics and scattering probability which can be expressed as
\begin{equation}\label{eq:zeta}
\zeta(\sigma, u_\chi, r') = \displaystyle\begin{dcases*}
\sum_S \dfrac{3}{2 R_\star} \int_0^1 dy \dfrac{y e^{-y \tau} \left( y \tau \right)^S}{S!} \, g_S \left(u_{\chi} \right); & \text{for BD and WD.} \\
\sum_i \int_0^{v_{\rm esc}(r')} dv R_i^- \left( w(r') \rightarrow v \right) \left| F_i(q) \right|^2; & \text{for MS stars.}
\end{dcases*}
\end{equation}
The index $S$ sums over the multiple scattering within a compact star. The optical depth of such a star is defined as $\tau = 3 \sigma \mathcal{N}_t/\left( 2 \pi R_\star^2 \right)$, where $R_\star$ is the radius of the star, $\mathcal{N}_t$ represents the total target constituents available for DM scattering and $\sigma$ is the DM-nuclei scattering cross-section. The term $g_S \left( u_\chi \right)$ denotes the probability of capture rate after $S$ scattering events which is extracted from \cite{Bramante:2017xlb,Dasgupta:2019juq,Ilie:2020vec}. On the other hand, the index $i$ accounts for the different chemical constituents as is relevant for a MS star like the Sun \cite{Vinyoles:2016djt}. In Eq. \eqref{eq:zeta}, the factor  $R_i^- \left( w(r') \rightarrow v \right)$ denotes the rate of DM scattering with the nuclei $i$ so that the velocity of the DM particle attains a lower velocity $v$ from the initial velocity $w(r)$ \cite{Busoni:2017mhe,Garani:2017jcj} and $\left| F_i(q) \right|^2$ is the nuclear form factor \cite{Garani:2017jcj}. The capture rates for brown dwarfs and main sequence stars have been computed using the \texttt{Asteria} package \cite{Leane:2023woh}, which provides a more accurate treatment of the optically thick region. For white dwarfs, we have determined the capture rates following the prescription described above and have verified that the corresponding cross-sections remain below the geometric limit.

Once the capture rate for a single star is known it is straight forward to generalise this for a distribution of stars as
\begin{equation}
\label{eq:C_tot}
C_{\rm tot} = \int_{r_1}^{r_2} dr \, 4 \pi r^2 \, n_\star(r) \, C_\star(r),
\end{equation}
where $n_\star(r)$ is the density of stars in our galaxy and has been adopted for MS \cite{Alexander:2008tq}, BD \cite{Leane:2021ihh}, WD\cite{Acevedo:2023xnu} and NS \cite{Generozov:2018niv} distributions between $r = \left[ 0.001 \right.$\,$-$\,$\left. 10 \right]\,$pc of the galactic center. The typical MS star is assumed to resemble the Sun. For other stellar types, the typical parameters are set  as follows: BD with a mass of $0.0378 \, M_{\odot}$ and a radius equivalent to that of Jupiter \cite{Leane:2021ihh}; WD with a mass of $0.49 \, M_{\odot}$ and a radius of $9390 \, \rm{km}$ \cite{Bell:2021fye}; and NS with a mass of $1.5 \, M_{\odot}$ and a radius of $10 \, \rm{km}$ \cite{Leane:2021ihh,Bose:2021yhz}. The DM density profiles range from cored to cuspy \cite{Gaskins:2016cha,Nesti:2013uwa}, and their effects on DM capture has been studied in \cite{Bose:2021yhz}. In this article, we consider the cuspy  generalized Navarro–Frenk–White (NFW) profile \cite{Moore:1999gc} and the profile is normalized to the local solar neighborhood density of $0.3 \, {\rm GeV/cm^3}$ \cite{Read:2014qva,ParticleDataGroup:2022pth,Salucci:2010qr}. We account for the variation of velocity dispersion as a function of distance from the galactic center extracted from \cite{Chemin2015TheIR} which is in consonance with the more detailed hydrodynamical simulations including non-circular gas flow \cite{Chemin2015TheIR,Li2021GasDI,Khrapov2021ModelingOS}. The distribution indicates lower velocities near the galactic center further aiding the DM capture rates.
%
%
%%%%%%%%%%%%%%%%%%%%%%%%%%%%%%%%%%%%%%%%%%%%%%%%%%%%%%%%%%%%%%%%%%%%%%%%%%%%%%%%%%%%%%%%%%%%
\section{Constraints on sub-GeV dark matter}
\label{sec:neutrinos_gal_center}
%%%%%%%%%%%%%%%%%%%%%%%%%%%%%%%%%%%%%%%%%%%%%%%%%%%%%%%%%%%%%%%%%%%%%%%%%%%%%%%%%%%%%%%%%%%%

After thermalisation, the DM particles that are captured inside a celestial body can evaporate and annihilate. However, for each of the stellar species under consideration, we focus on the region of parameter space where the evaporation effects  are  numerically insignificant \cite{Garani:2021feo}. When DM particles directly annihilate into SM states, neutrinos are generated within the star either via primary or secondary annihilation processes. The differential neutrino flux at the Earth can be expressed as
%
%
%%%%%%%%%%%%%%%%%%%%%%%%%%%%%%%%%%%%%%%%%%%%%%%%%%%%%%%%%%%%%%%%%%%%%%%%%%%%%%%%%%%%%%%%%%%%
\begin{figure}[t]
\centering
\includegraphics[width=0.925\linewidth]{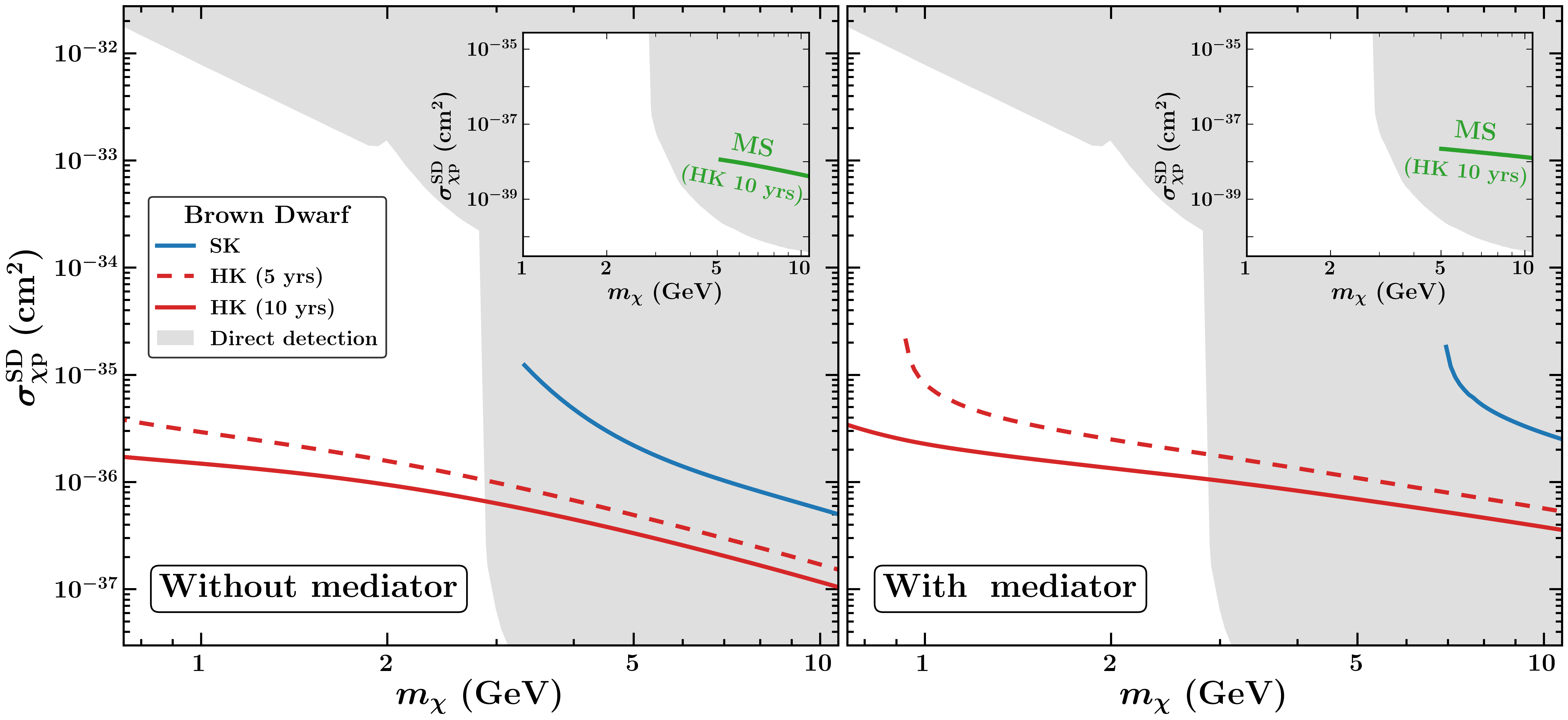}
\caption{Projected limits on the SD scattering cross-section of DM with nucleons are derived for two scenarios: direct annihilation of captured DM particles to neutrinos (left) and annihilation into long-lived mediators (right) considering the DM capture in BD distribution. The blue solid line represents the sensitivity limits derived from the reach of Super-Kamiokande, while the red dashed and solid lines correspond to the projected sensitivity of Hyper-Kamiokande, assuming operational timescales of $5\,$yrs and $10\,$yrs respectively. The sub-dominant constraint from the MS distribution is shown in the inset by the green solid line. The gray shaded region represents the direct detection limits from CDMSlite \cite{SuperCDMS:2017nns}, PICO-60 \cite{PICO:2019vsc}, and XENON1T \cite{XENON:2019zpr}.}
\label{fig:SDp_limit_both_Med_LowMass}
\end{figure}
%%%%%%%%%%%%%%%%%%%%%%%%%%%%%%%%%%%%%%%%%%%%%%%%%%%%%%%%%%%%%%%%%%%%%%%%%%%%%%%%%%%%%%%%%%%%
%
%
%
\begin{equation}
\label{eq:diff_flux_wo_Med}
\left. E_\nu^2 \dfrac{d \phi_\nu}{d E_\nu} \right|_{\rm Earth} = \dfrac{\Gamma_{\rm ann}}{4 \pi D_{\rm GC} ^2} \times \left. E_\nu^2 \dfrac{d N_\nu}{d E_\nu} \right|_{\rm Prop.},
\end{equation}
where the annihilation rate becomes $\Gamma_{\rm ann} = C_{\rm tot}/2$ under the equilibrium condition and $D_{\rm GC}$ is the distance of the Earth from the galactic center. The $\left. d N_\nu/d E_\nu \right|_{\rm Prop. }$ has been obtained by first generating the spectrum inside the core using ${\chi \rm{aro} \nu}$ \cite{Liu:2020ckq} and then considering the propagation of the neutrinos within the stellar interior, galactic halo, Earth atmosphere and crust using \texttt{nuSQuIDS} \cite{Arguelles:2021twb}. For the results presented here, we have assumed a direct annihilation of DM to the neutrinos democratically distributed in all flavors. At the Earth based neutrino detectors, the incident neutrinos can be identified through either track-like events generated by charge-current interactions of $\nu_\mu$ or cascade events originating from $\nu_e$ and $\nu_\tau$. We focus on muon events due to their superior angular resolution crucial to identify the galactic center in the sky. The differential muon events resulting from the muon neutrino flux within the detector are given by \cite{Kistler:2006hp}
\begin{equation}
\label{eq:dN_mu-dE_mu}
\dfrac{d N_\mu}{d E_\mu} = N_A \rho V T \dfrac{1}{1 - y} \left[ \dfrac{d \phi_{\nu_\mu} (E_\nu)}{d E_{\nu_\mu}} \sigma_{\rm CC}(E_{\nu_\mu}) \right]_{E_\nu = \dfrac{E_\mu}{1-y}},
\end{equation}
where $\rho$, $V$, $T$ are the density, effective volume and exposure time of the detector respectively and $N_A$ is the Avogadro number. The charge current induced $\nu_\mu$ cross-section $(\sigma_{\rm CC})$ has been adopted from \cite{Bell:2020rkw}. The average muon energy $\left\langle E_\mu \right\rangle$ is related to the neutrino energy $E_\nu$ by $\left\langle E_\mu \right\rangle = E_\mu = E_\nu (1-y)$. The primary source contributing to the background for this neutrino search is the atmospheric neutrino background, for which we have utilised the averaged neutrino flux data from \cite{Honda:2015fha}. The background estimate adopted in our analysis is comparable with the direct measurements from Super-Kamiokande \cite{Super-Kamiokande:2015qek} and provides a conservative bound on the parameter space. The other source of neutrino background arising from cosmic ray interactions in the solar atmosphere \cite{Ng:2017aur,Edsjo:2017kjk} remains below the atmospheric neutrino background and can be effectively reduced in directional searches around the galactic center. Unfortunately, the angular resolution of Super-Kamiokande is relatively limited for these low energetic neutrinos. To account for this low angular resolution, we have accounted for signal and background from a larger angular region surrounding the galactic center, approximated as $\delta \theta \simeq 30^{\circ} \left( {\rm GeV}/E_\nu \right)^{0.5}$ \cite{Yuksel:2007ac}. Although Hyper-Kamiokande is expected to have an improved angular resolution, we have set the angular resolution to be similar to that of Super-Kamiokande to conservatively obtain competitive limits.

To assess the sensitivity, we computed the number of signal events and the total background events within the energy bin $\left[ \text{Max}(E_{\text{thres}}, m_\chi/5), m_\chi \right]$, following \cite{Mandal:2009yk,Dasgupta:2012bd}. The $\sim 10\%$ uncertainty in the bin choice that may arise from the energy resolution of the detector does not have a numerically significant impact on our results for most part of the parameter space. We have considered the fiducial volume of Super-Kamiokande to be $22.5\,$kton with a total exposure time of $6050.3\,$days \cite{Super-Kamiokande:2022ncz}. For the Hyper-Kamiokande detector, we consider a fiducial volume of $187\,$kton \cite{Bell:2020rkw}, with projections based on $5\,$yrs and $10\,$yrs of operational time. In Fig. \ref{fig:Event_wo_Med_WD}, we present a benchmark plot for the background and signal event rates from WD distribution around galactic center, calculated using Eq. \eqref{eq:dN_mu-dE_mu} for DM mass of $5 \,$GeV and SI scattering cross-section $\sigma_{\chi p}^\text{SI} = 10^{-40} \, {\rm cm^2}$. The constraints on the spin-dependent (SD) DM scattering cross-section with protons is shown in Fig. \ref{fig:SDp_limit_both_Med_LowMass}. In the left panel of Fig. \ref{fig:SDp_limit_both_Med_LowMass}, we illustrate the limits obtained for captured  DM annihilation  to neutrinos within the star. This is contrasted with the scenario where the captured DM predominantly annihilate through the long-lived mediators in the right panel. The details of the analysis for the long-lived mediator scenario is discussed in \ref{app:other_limits}. The gray shaded region in Fig. \ref{fig:SDp_limit_both_Med_LowMass} represents the direct detection limits obtained from CDMSlite \cite{SuperCDMS:2017nns}, PICO-60 \cite{PICO:2019vsc}, and the Migdal limits from XENON1T \cite{XENON:2019zpr}. In Fig. \ref{fig:SDp_limit_both_Med_LowMass}, the constraints primarily consider the BD distribution which offers the stronger limits than the direct detection. The blue solid line corresponds to the sensitivity of the Super-Kamiokande detector while the red dashed and solid lines represent the projected reach of the Hyper-Kamiokande detector for $5\,$yrs and $10\,$yrs of exposure time respectively. The subdominant constraints from the MS population are illustrated by the green line in the inset of Fig. \ref{fig:SDp_limit_both_Med_LowMass}.
%
%
%%%%%%%%%%%%%%%%%%%%%%%%%%%%%%%%%%%%%%%%%%%%%%%%%%%%%%%%%%%%%%%%%%%%%%%%%%%%%%%%%%%%%%%%%%%%
\begin{figure}[t]
\centering
\includegraphics[width=0.85\linewidth]{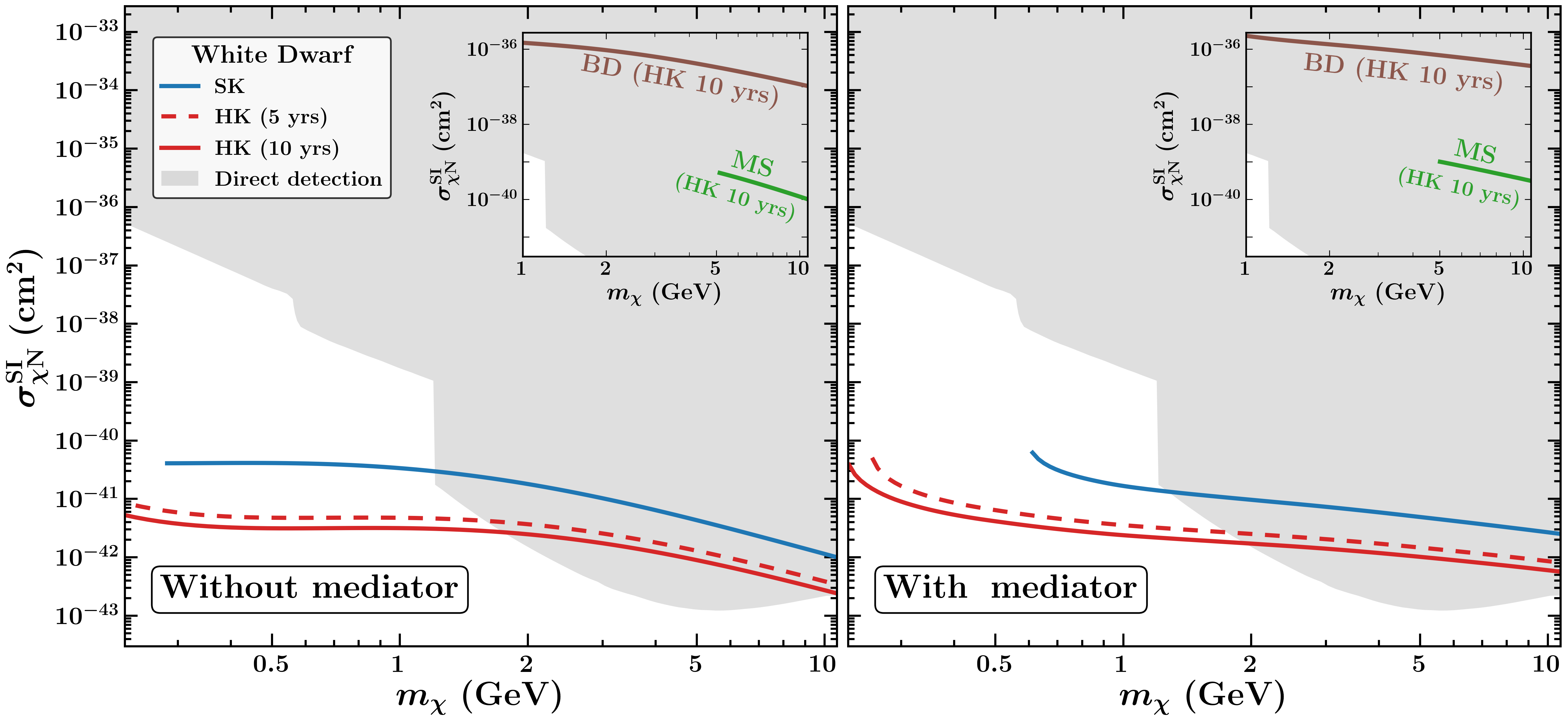}
\caption{Same as Fig. \ref{fig:SDp_limit_both_Med_LowMass} for SI scattering cross-section considering DM capture in WD distribution. In the inset, additional sub-dominant limits derived from the BD and MS distributions are shown with brown and green lines respectively. The direct detection limits from DarkSide-50 \cite{DarkSide-50:2022qzh} and Migdal-XENON1T \cite{XENON:2019zpr} are depicted by the gray shaded region.}
\label{fig:SI_limit_both_Med_LowMass}
\end{figure}
%%%%%%%%%%%%%%%%%%%%%%%%%%%%%%%%%%%%%%%%%%%%%%%%%%%%%%%%%%%%%%%%%%%%%%%%%%%%%%%%%%%%%%%%%%%%
%
%

In Fig. \ref{fig:SI_limit_both_Med_LowMass}, we depict the limits on the spin-independent (SI) DM scattering cross-section with nucleons considering the neutrinos arising from DM capture in WD distribution. Since WDs are primarily composed of carbon nuclei, only spin-independent interaction with DM is feasible. As a result, the corresponding limits are not shown in Fig. \ref{fig:SDp_limit_both_Med_LowMass}. Similar to Fig. \ref{fig:SDp_limit_both_Med_LowMass}, the blue solid line indicates the sensitivity of the Super-Kamiokande detector, while the red dashed and solid lines represent the projected reach of the Hyper-Kamiokande detector for $5\,$yrs and $10\,$yrs of operational time respectively in Fig. \ref{fig:SI_limit_both_Med_LowMass}. However, these results correspond to neutrinos originating from dark matter capture within the WD distribution at the galactic center. In Fig. \ref{fig:SI_limit_both_Med_LowMass}, the gray shaded region representing the direct detection bounds, incorporates limits from DarkSide-50 \cite{DarkSide-50:2022qzh,GrillidiCortona:2020owp,DarkSide:2022dhx,DarkSide-50:2023fcw} and the Migdal lines from XENON1T \cite{XENON:2019zpr}. For SI interaction, both the BD and MS distributions yield sub-leading constraints compared to those from direct detection experiments which offer enhanced sensitivity. These constraints from BD and MS populations are depicted in the inset of Fig. \ref{fig:SI_limit_both_Med_LowMass} represented by the brown and green lines respectively.

In the absence of a mediator, sub-GeV neutrinos are unable to escape the neutron star which explains the lack of constraints. However, in the case of a dark mediator, neutrinos could be produced via mediator decay, but the resulting flux remains undetectable by Earth-based detectors even with the geometric capture rate. This is primarily due to the relatively lower population of neutron stars compared to other stellar populations. For both SD and SI scenarios, we find that the naturally expected neutrino signals from the captured DM improves the present limits from direct detection in $200\,$MeV\,$-$\,$2\,$GeV mass scales. For the SD interaction, the BD distribution provides the most stringent bounds while for SI case, the WD distribution is the most optimistic. Inspite of its population, the MS stars provide sub-leading constraints primarily due to its low density. Curiously we conclude that the observable neutrino signals and hence constraints on the low mass sub-GeV scale captured DM are expected from galactic distribution of compact dwarf stars. Other dedicated limits from the long-lived mediator scenarios can be found in appendix \ref{app:other_limits}.
%
%
%%%%%%%%%%%%%%%%%%%%%%%%%%%%%%%%%%%%%%%%%%%%%%%%%%%%%%%%%%%%%%%%%%%%%%%%%%%%%%%%%%%%%%%%%%%%
\section{Conclusion}
\label{sec:conclusion}
%%%%%%%%%%%%%%%%%%%%%%%%%%%%%%%%%%%%%%%%%%%%%%%%%%%%%%%%%%%%%%%%%%%%%%%%%%%%%%%%%%%%%%%%%%%%

Captured dark matter naturally settles near the core of the stars and dissipates its energy through annihilation to visible sector states. The most expected signatures include a thermal heating of the stars that has been widely discussed in the context of the low brightness neutron stars in the literature. In this article, we present the complimentary signature  of low energy neutrinos from the stellar cores produced by annihilation of captured dark matter that is expected in this scenario without any additional assumptions. We find that the dense stellar interiors together with the rapid rise of neutrino scattering cross-section with energy restricts the possibility of observing such signals. Interestingly, a viable window is explored for the DM mass range between $200\,$MeV\,$-$\,$2\,$GeV where neutrinos can emerge from the stellar core for a distribution of MS, WD and BD stars near the galactic center and produce an observable flux at neutrino observatories like the Super-Kamiokande or the upcoming Hyper-Kamiokande. A synergy of a cuspy DM density profile with lower halo velocities and an optimistic distribution of stars near the galactic center can push the projected limits below the existing direct detection bounds in the sub-GeV mass domain. The sensitivity limits for upcoming Hyper-Kamiokande have been determined assuming a larger angular resolution comparable to that of Super-Kamiokande. With expected better angular resolution for the galactic center and improved background discrimination at future observatories, the limits are expected to improve significantly in a region where the direct detection experiments hit their resolution limits.
%
%
%%%%%%%%%%%%%%%%%%%%%%%%%%%%%%%%%%%%%%%%%%%%%%%%%%%%%%%%%%%%%%%%%%%%%%%%%%%%%%%%%%%%%%%%%%%%
\begin{figure}[t]
\centering
\includegraphics[width=0.5\linewidth]{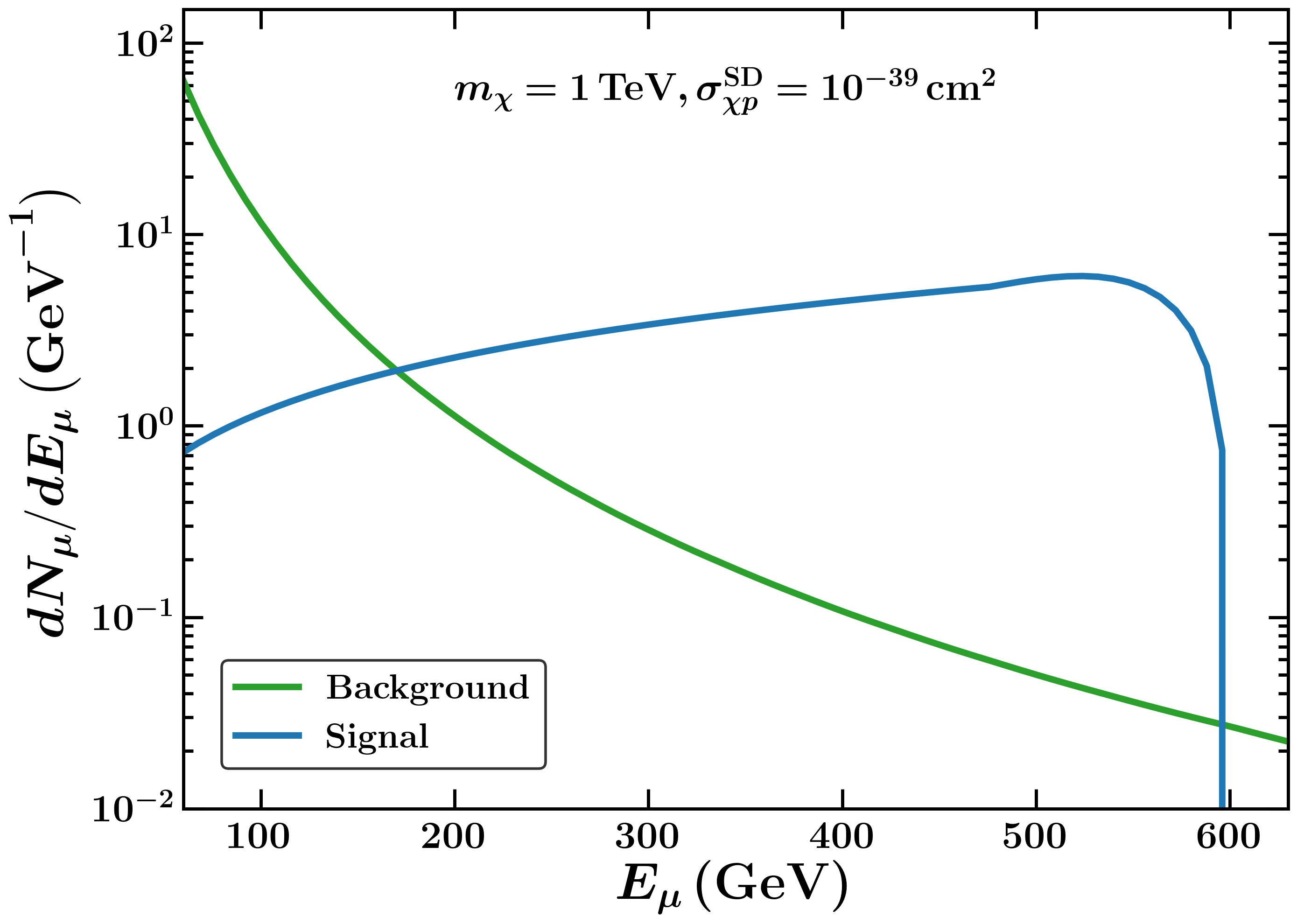}
\caption{Same as Fig. \ref{fig:Event_wo_Med_WD} but for the mediator framework for DM mass of $1\,$TeV with the MS distribution, considering a spin-dependent (SD) scattering cross-section of \(10^{-39} \, \text{cm}^2\).}
\label{fig:Event_wt_Med_MS}
\end{figure}
%%%%%%%%%%%%%%%%%%%%%%%%%%%%%%%%%%%%%%%%%%%%%%%%%%%%%%%%%%%%%%%%%%%%%%%%%%%%%%%%%%%%%%%%%%%%
%
%
%%%%%%%%%%%%%%%%%%%%%%%%%%%%%%%%%%%%%%%%%%%%%%%%%%%%%%%%%%%%%%%%%%%%%%%%%%%%%%%%%%%%%%%%%%%%
\paragraph*{Acknowledgments\,:} We thank Carlos Argüelles-Delgado, Qinrui Liu and Tarak Nath Maity for helpful discussions. DB acknowledges MHRD, Government of India for fellowship.
%%%%%%%%%%%%%%%%%%%%%%%%%%%%%%%%%%%%%%%%%%%%%%%%%%%%%%%%%%%%%%%%%%%%%%%%%%%%%%%%%%%%%%%%%%%%
%
%
\appendix
%
%
%
%
%%%%%%%%%%%%%%%%%%%%%%%%%%%%%%%%%%%%%%%%%%%%%%%%%%%%%%%%%%%%%%%%%%%%%%%%%%%%%%%%%%%%%%%%%%%%
\section{Long-lived mediator scenario}
\label{app:other_limits}
%%%%%%%%%%%%%%%%%%%%%%%%%%%%%%%%%%%%%%%%%%%%%%%%%%%%%%%%%%%%%%%%%%%%%%%%%%%%%%%%%%%%%%%%%%%%

As depicted in Fig. \ref{fig:Neutrino_flux_wo_Med}, when DM annihilates to neutrinos within the stellar core, the high energy neutrinos are hindered from emerging from the stellar atmosphere due to scattering. However, DM may interact with SM particles via long-lived mediators \cite{Pospelov:2007mp,Pospelov:2008jd,Batell:2009zp,Chen:2009ab} that can escape the star with minimal attenuation making the capture signals detectable at Earth based observatories. With these long-lived mediators, one can expect signature of captured DM in the form of observable gamma rays \cite{HAWC:2018szf,Bell:2021pyy,Leane:2021ihh,Leane:2021tjj,Bose:2021yhz,Bose:2021cou,Acevedo:2023xnu,Chen:2023fgr,Gustafson:2023hvm,Linden:2024uph,Leane:2024bvh}, neutrinos \cite{Bell:2011sn,Leane:2017vag,Niblaeus:2019gjk,Nguyen:2022zwb,Chu:2024gpe}, and charged particles \cite{Feng:2016ijc,Smolinsky:2017fvb,Li:2022wix}.

The differential neutrino flux from the captured DM annihilation through long-lived mediators is schematically given by \cite{Leane:2024bvh}
\begin{gather}
\label{eq:diff_flux_wt_Med}
\left. E_\nu^2 \dfrac{d \phi_\nu}{d E_\nu} \right|_{\rm w/t \ Med} = \dfrac{\Gamma_{\rm ann}}{4 \pi D_{\rm GC}^2} \times {\rm Br \left( Y \rightarrow SM \, \bar{SM} \right)} \times \left. \dfrac{1}{3} E_\nu^2 \dfrac{d N_\nu}{d E_\nu} \right|_{\chi \bar{\chi} \rightarrow \rm{Y Y} \rightarrow {2 \left( \rm SM \, \overline{SM} \right) } } \times \mathbb{P}_{\rm surv} \, ,
\end{gather}
where ${\rm Br \left( Y \rightarrow SM \, \bar{SM} \right)}$ represents the branching ratio of the mediator decaying to a SM final state. The term $\left. d N_\nu/d E_\nu \right|_{\chi \bar{\chi} \rightarrow \rm{Y Y} \rightarrow {2 \left( \rm SM \, \overline{SM} \right) } }$ denotes the neutrino spectrum resulting from DM annihilation via mediator \cite{Elor:2015bho}. We have assumed a direct decay of the mediators to a pair of neutrinos making the spectra relatively independent of the mediator mass. However, our obtained limits are directly applicable to DM models where the long-lived mediator mass is lighter than MeV scale to prevent annihilation into other channels. This may be contrasted with massive promptly decaying mediators discussed in Sec. \ref{sec:neutrinos_from_stars} which are expected to be $\mathcal{O}(100)\,$GeV to reproduce the correct relic abundance \cite{Lin:2022hnt}. Following the discussion in \cite{Leane:2024bvh}, we have assumed the survival probability $\mathbb{P}_{\rm surv} \sim 1$ which holds true for most of the mediator parameter space given the large distance between the galactic center and the Earth. Given the considerable distance traveled by neutrinos, we can reasonably neglect the oscillation effects and assume the flux has been averaged over all the neutrino flavours. In Figs. \ref{fig:SDp_limit_both_Med_LowMass} and \ref{fig:SI_limit_both_Med_LowMass}, we have emphasized the constraints achievable with Super-Kamiokande and Hyper-Kamiokande, contrasting them with the scenario without any long-lived mediator for SD and SI interactions, respectively.
%
%
%%%%%%%%%%%%%%%%%%%%%%%%%%%%%%%%%%%%%%%%%%%%%%%%%%%%%%%%%%%%%%%%%%%%%%%%%%%%%%%%%%%%%%%%%%%%
\begin{figure*}[t]
\centering
\begin{subfigure}{0.425\textwidth}
\centering
\includegraphics[width=0.975\linewidth]{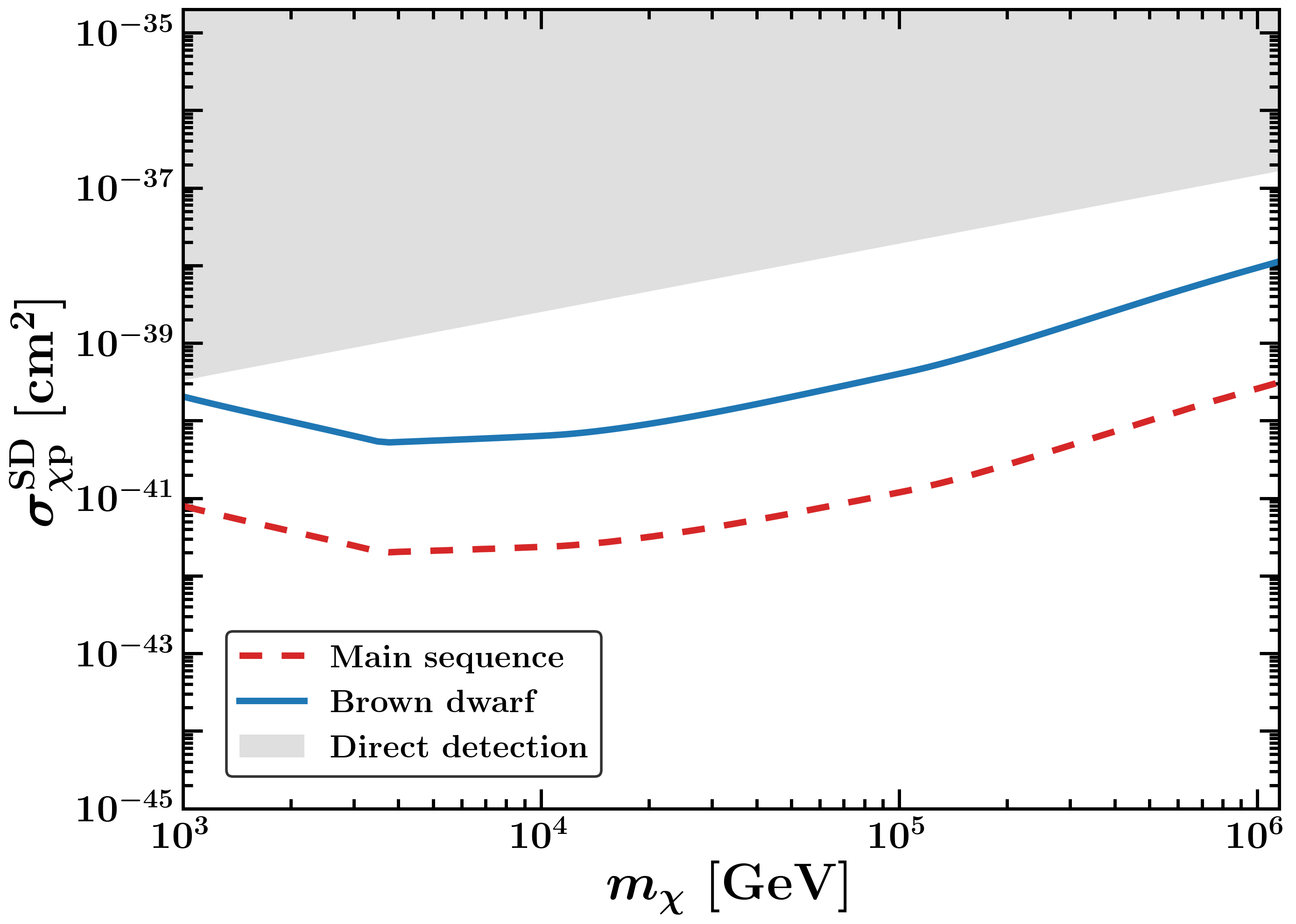}
\caption{}
\label{sf:SDp_limit_wt_Med_HighMass}
\end{subfigure}
\begin{subfigure}{0.425\textwidth}
\centering
\includegraphics[width=0.975\linewidth]{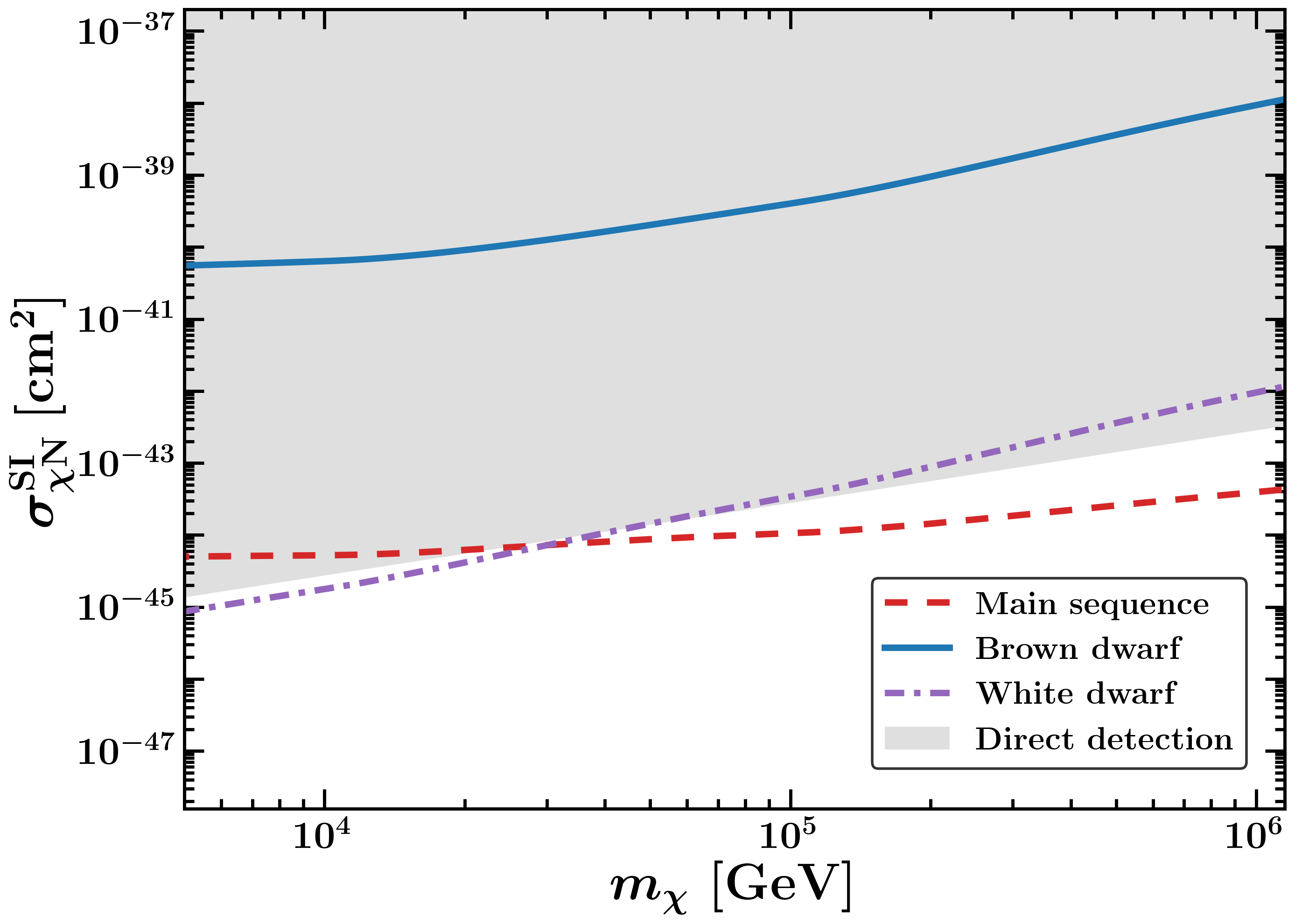}
\caption{}
\label{sf:SI_limit_wt_Med_HighMass}
\end{subfigure}
\caption{Projected limits on the DM parameter space for long-lived mediators models from KM3NeT. The red (dashed), blue (solid), and purple (dot-dashed) lines represent the MS, BD, and WD distributions respectively, while the gray shaded region denotes the envelope of direct detection limits.}
\label{fig:dark_mediator_limits}
\end{figure*}
%%%%%%%%%%%%%%%%%%%%%%%%%%%%%%%%%%%%%%%%%%%%%%%%%%%%%%%%%%%%%%%%%%%%%%%%%%%%%%%%%%%%%%%%%%%
%
%

In Fig. \ref{fig:dark_mediator_limits}, we emphasize the sensitivities of neutrino searches applicable to the long-lived mediator scenario for WD, BD and MS populations while the corresponding bounds from the distributions of NS can be found in \cite{Bose:2021yhz}. We depict the reach of a gigaton detector like KM3NeT \cite{KM3Net:2016zxf} or IceCube \cite{IceCube:2016yoy} for high-mass DM. Given the geographical location of KM3NeT, it is particularly suitable for detecting neutrinos originating from the galactic center. Our results focus on the KM3NeT detector, where the galactic center remains approximately below the horizon for about $37\%$ of the total exposure time \cite{Ng:2020ghe}. We have assumed an operational time of $10\,$yrs for the KM3NeT detector to ensure a sufficient number of measurable events. The $90\%$ confidence level bounds for high DM masses are derived through the same analysis, utilizing energy bins $ \left[ \text{Max}(E_{\text{thres}}, m_\chi/5), m_\chi \right]$ with an opening angle $\theta_{\nu \mu} = 0.7^{\circ} \left(10^3 \, \rm{GeV}/ E_\nu \right)^{0.6}$ \cite{KM3NeT2008}. In Fig. \ref{fig:Event_wt_Med_MS}, we present the differential event rates for background (green) and signal (blue) for a DM mass of $1\,$TeV with a SD scattering cross-section of \(10^{-39} \, \text{cm}^2\), obtained by considering the MS distribution of stars in the galactic center. To derive the limits in the high mass regime, we have imposed a requirement of at least $10$ signal events in the KM3NeT detector within the operational time. Figs. \ref{sf:SDp_limit_wt_Med_HighMass} and \ref{sf:SI_limit_wt_Med_HighMass} present the projected SD and SI limits respectively derived using the KM3NeT detector across various stellar distributions. As expected, the WD distribution does not influence the SD constraints due to its carbon-dominated composition, while similar results for NS distribution are detailed in \cite{Bose:2021yhz}.  The direct detection limits are obtained from PICO-60 \cite{PICO:2019vsc} and LUX-ZEPLIN \cite{LZ:2022lsv} for SD and SI interactions respectively. Fig. \ref{fig:dark_mediator_limits} illustrates that prospective limits from neutrino searches originating from DM captured within stellar distributions impose more stringent constraints than direct detection limits.
%
%
%%%%%%%%%%%%%%%%%%%%%%%%%%%%%%%%%%%%%%%%%%%%%%%%%%%%%%%%%%%%%%%%%%%%
%
%
%%%%%%%%%%%%%%%%%%%% References %%%%%%%%%%%%%%%%%%%%%%%%%%%%%%%%%%%%

\let\OLDthebibliography\thebibliography
\renewcommand\thebibliography[1]{
  \OLDthebibliography{#1}
  \setlength{\parskip}{0pt}
  \setlength{\itemsep}{0pt plus 0.1ex}
}
\bibliographystyle{JHEP}
\bibliography{neutrinos_from_stellar_dist.bib}
%%%%%%%%%%%%%%%%%%%%%%%%%%%%%%%%%%%%%%%%%%%%%%%%%%%%%%%%%%%%%%%%%%%%
%
%
%
%
%
\end{document}